\documentclass{article}                     

\usepackage{graphicx}
\usepackage{subfigure}

\begin{document}

\begin{center}
{\Large  \bf Startup of the High-Intensity Ultracold \newline 
Neutron Source at the Paul Scherrer Institute}\\

\vspace*{10mm}
{{\bf Bernhard Lauss}\\
{\small on~behalf~of~the~PSI~UCN~Project~Team
\footnote{The members of the PSI UCN Project Team are listed 
at \texttt{http://ucn.web.psi.ch}.}}}\\
\vspace*{10mm}
{Paul Scherrer Institute, CH-5232 Villigen-PSI, Switzerland}

\end{center}


\begin{abstract}

Ultracold neutrons (UCN) can be stored in suitable bottles 
and observed for several hundreds of seconds.
Therefore UCN can be used to study in detail
the fundamental properties of the neutron.
A new user facility providing ultracold neutrons for fundamental physics
research has been constructed at the Paul Scherrer Institute, the PSI UCN source. 
Assembly of the facility
finished in December 2010 with the first production 
of ultracold neutrons. Operation approval was received in June 2011. 
We give an overview of the source and the status at startup.


\end{abstract}

\section{Introduction}

The fundamental properties of the neutron can serve as
a window into the first minutes of the universe.
The neutron lifetime is an important parameter in
understanding the abundance of light elements created in 
the big-bang nucleo-synthesis \cite{pdg}.
Finding a neutron electric dipole moment (nEDM)\cite{golub}
would hint at a new CP violating process necessary to 
understand the asymmetry of matter over antimatter in our universe.
nEDM searches 
(e.g.\cite{kirch,baker})  
probing physics beyond the Standard Model 
are presently statistics limited and are the main
driving force behind the proposals of
several new high-intensity ultracold neutron sources
around the globe \cite{NuclPhysNews}.
Moreover, stored neutrons allow to 
search for extra forces not included in the 
Standard Model of particle physics 
or search for 
modifications of Newtonian gravity in the sub-millimeter range
predicted by string theories \cite{abele}.

\begin{figure}
\centering
  \includegraphics[height=.42\textheight]{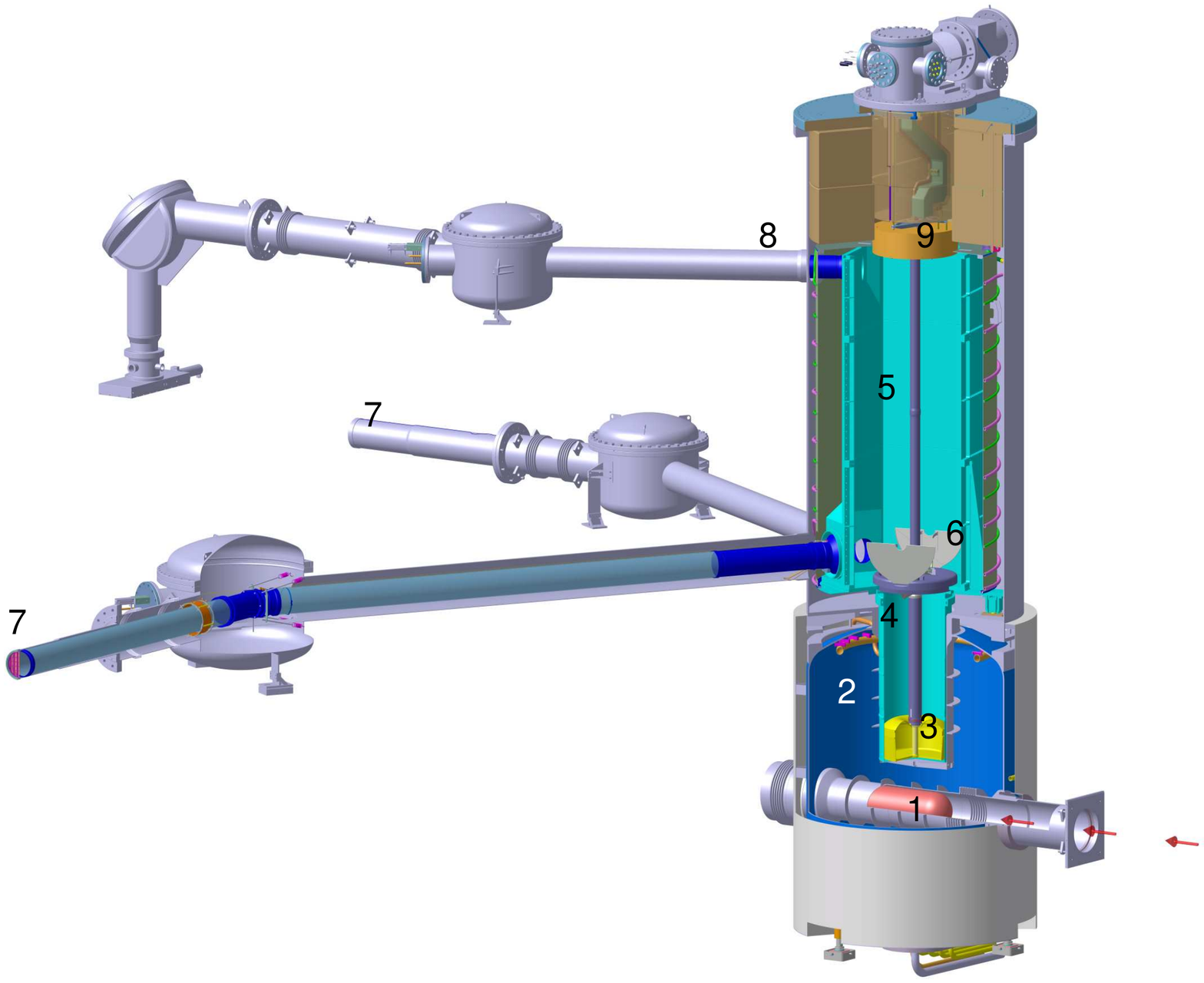}
  \caption{
  Cut view of the neutron production, storage and guiding system of the PSI UCN source
  inside a 7~m high vacuum tank.
The indicated components are described in the text.
  }
\vspace*{-5mm}
\label{source}
\end{figure}

\section{The ultracold neutron source}

Neutrons with kinetic energies below $\sim$350~neV, 
corresponding to a few milli-Kelvin, are termed ultracold neutrons (UCN).
The material optical potential of certain materials (e.g. Ni, Be, steel, 
diamond-like carbon / DLC)
is high enough 
-- due to their high density and large neutron bound-coherent scattering length --
that UCN undergo total reflection under all angles of incidence \cite{golub}.
Hence, UCN can be stored in material bottles for several hundreds of
seconds. 
They can also be contained via gravity, with an 
energy change of 100~neV per meter and 
they can be manipulated via their magnetic moment, where a 1 Tesla
field change corresponds to a change of 60~neV in potential energy.

Over the last years an ultracold neutron source based
on accelerator driven spallation neutron production
has been constructed at the Paul Scherrer Institute (PSI), Switzerland
\cite{blau,laussinpc}.

\begin{figure}
\centering
  \includegraphics[height=.35\textheight]{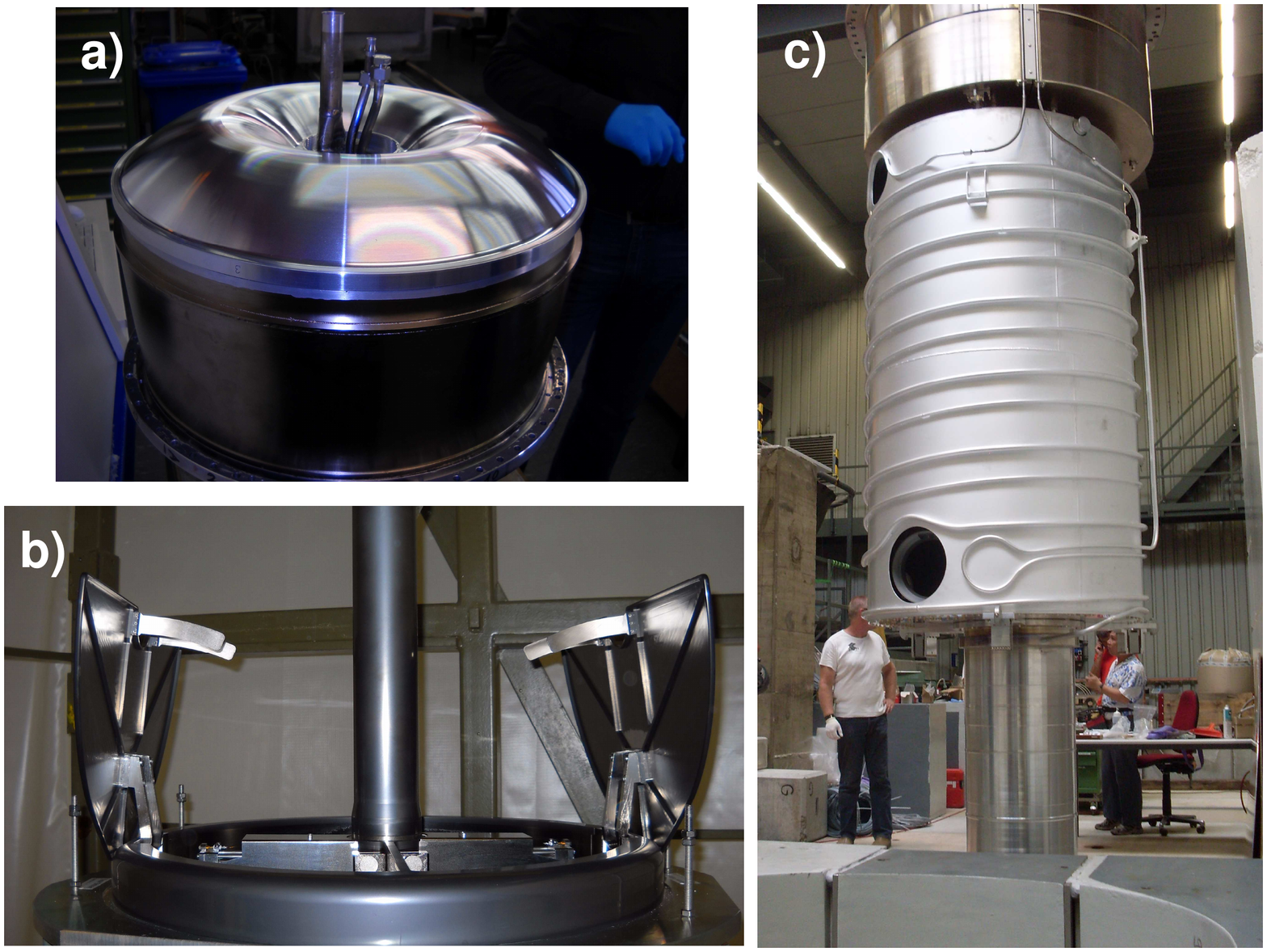}
  \caption{Parts of the UCN source during assembly.
  a) The container for the solid deuterium crystal. The donut shaped
  top-lid, necessary for pressure safety, is only 0.5~mm thick to 
  allow for a high UCN transmission.\newline
  b) The storage volume shutter during operation testing. The 2
  aluminum jaws visible on each flap are used for stopping the closing
  flaps via an eddy-current brake working at 80~K.\newline
  c) The UCN storage volume before insertion in the vacuum tank. 
  Visible is the vertical neutron guide on the bottom, the radiation shield
  surrounding the UCN storage volume with the 80~K cooling tubes and the opening for
  the neutron guides. The steel shielding is on top.
  }
\vspace*{-5mm}
\label{fig-parts}
\end{figure}

The method, experimentally pioneered at PNPI \cite{serebrov}
and Los Alamos National Lab \cite{lanl-ucn}, 
is based on
i) neutron production via proton induced spallation on lead \cite{wohl};
ii) superthermal UCN production in solid ortho-deuterium (s$D_2$) \cite{golub};
and iii) intermediate UCN storage and distribution of UCN on demand to experiments.
The main experimental components covering the neutron production and 
transport are shown in Fig.\ref{source}. 
PSI's 590~MeV proton beam with up to 2.4~mA beam 
current is impinging on a heavy-water-cooled 
canneloni-type lead spallation target (see (1) in Fig.\ref{source}) \cite{wohl}. 
UCN operation is limited by radiation safety requirements to an 
integrated proton flux of 20~$\mu$A/hour. 
This means that a 2~mA proton beam kick of 8~s 
is followed by a minimum waiting period of 800~s.
Spallation neutrons are thermalized in the surrounding 
heavy water tank (2). 
Close to the spallation target inside a AlMg3 container
(No.(3) in Fig.\ref{source} and Fig.\ref{fig-parts}a)
a pure $\sim$30 liter ortho-deuterium crystal is kept at a temperature of 5~K. 
It serves to moderate neutrons first to be cold and finally to be ultracold neutrons
via downscattering \cite{golub,NuclPhysNews,malgosia}.

The UCN production rate strongly depends on 
the deuterium temperature and spin \cite{NuclPhysNews,liu,malgosia}
favoring solid ortho-deuterium as production medium. 
Some UCN can then be emitted into the vacuum on
top of the crystal, where they get a 102~neV boost from the s$D_2$ material potential
at the crystal surface.
Via a vertical UCN guide (4) 
(shown in Fig.\ref{fig-parts}c)
they can reach the UCN storage volume (5) 
coated with diamond-like carbon which has a high material optical 
potential of about 235~neV and a very low neutron loss probability.

\begin{figure}
\centering
\includegraphics[height=.41 \textheight]{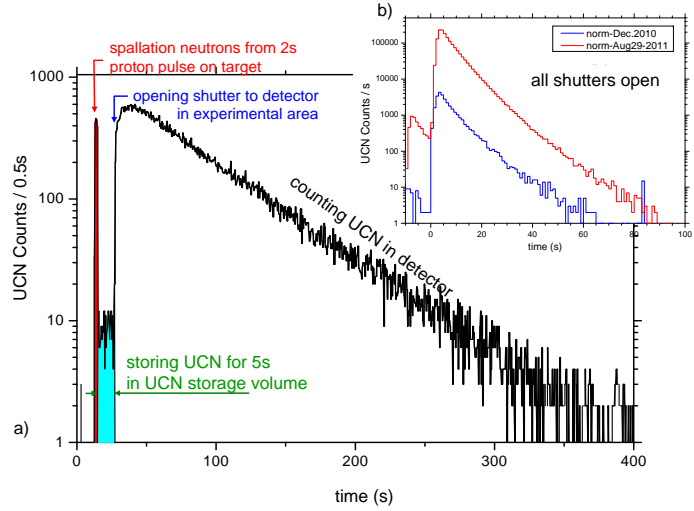}
\caption{
a) 
First UCN in December 2010: 
UCN counts observed in a Cascade-U detector located at the end of
the UCN guide in area West, 
plotted versus time (seconds).
The structure in the count rate reflects the initial 
proton beam kick onto the spallation target of 2~s which 
produce a neutron background in the detector. 
After the beam kick the UCN storage
volume is closed and the UCN are stored for 5~s.  
A few percent of the UCN leak during that time 
through the guide UCN shutter.
After 5~s the guide shutter towards the experimental area is opened
and the UCN can arrive at the detector. 
The observed count rate then reflects the exponential 
emptying of the UCN storage volume into the detector.
\newline
b) 
UCN counts versus time after beam kick 
observed on guide West-1 for `normalized pulses',
i.e. beam kicks with 2~s length and all shutters, including
the main UCN storage volume shutter remaining open all the time.
Hence the emptying of the storage volume is determined by
the large opening towards the deuterium container. 
The given time distributions are for the best beam kick
observed in Dec. 2010 and the startup kicks
in Aug. 2011. A factor 50 improvement is obvious.
}
\vspace*{-5mm}
\label{fig-ucn}
\end{figure}

After the beam kick the main UCN shutter ( No.(6) and Fig.\ref{fig-parts}b)
is rapidly closed and
the produced UCN are stored inside the volume.
Three UCN guides (7) and (8) lead through the radiation shielding to experimental areas. 
Shutters can open and close these guides at the storage volume exits. 
In order to provide optimal UCN transmission through more than 8~m of tubes
penetrating the radiation shielding,
most of the guides are made of 180~mm inner diameter 
Duran\textsuperscript{\textregistered}
glass tubes with 
approximately 1~nm surface roughness and 
sputter-coated with 500~nm thick NiMo on the inside.
A 80~cm long polished steel guide connects the glass guides at room temperature to 
the UCN storage volume which is operating at 80~K.
Radiation protection requires 
30$^\circ$ bends (made from polished stainless steel)
to prevent direct sight onto the storage volume.
On top of the UCN storage volume a large, activated carbon loaded cryo-pump is 
the coldest spot
of the source vacuum in order to efficiently capture residual gas atoms and 
prevent them from condensing on the s$D_2$ container
or anywhere else inside the storage volume.

Several important components are not shown on the figure. 
Most prominently the cryo-system necessary for the 
production of the ortho-deuterium crystal.
The present s$D_2$ production procedure \cite{alexander} uses
30~m$^3$ pure $D_2$ gas.
The gas is frozen into the `condenser' volume. 
Then it is liquefied and transported to the nearby 
`para-otho-converter' filled with a paramagnetic chromium-oxide 
(Oxisorb\textsuperscript{\textregistered})\cite{bodek} 
and left boiling at 19~K for several hours.
Raman spectroscopy of the rotational transitions in $D_2$ on
extracted gas samples
shows an ortho-deuterium concentration of 97$\pm$2$\%$ as expected
for equilibrium conditions at 19~K.
Finally, the ortho-$D_2$ is slowly frozen inside the crystal container. 
This process takes several days and, being a very important 
part of the source setup, still needs to be optimized. 

Final assembly of all parts and commissioning of the PSI
source finished in 2010. December 16, 2010 saw the first
UCN production during testing the radiological and cryogenic safety.
Fig.\ref{fig-ucn}a shows one of the first production fills
with a UCN count-rate structure as expected.
Beam kicks of up to 8~s length and full beam power were tested.

Operation approval of the Swiss Federal Authorities was
received on June 27 and
the source operation started in August, 2011.
The source performance is continuously improving 
since then with presently already a factor 50 
higher yield in comparison to Dec. 2010
as demonstrated in Fig.\ref{fig-ucn}b.
This can be mainly attributed to increasing the 
ortho-deuterium content from 66$\%$ to $\sim$97$\%$
and a different freeze-out technique.
A factor of about 30 lies ahead to full design intensity.

The experimental search for a neutron EDM is starting in 
area South with an installed apparatus \cite{kirch}. 
The UCN user facility at PSI is open for experiment proposals.

\paragraph{Acknowledgements}

Cordial thanks to the more than two hundred colleagues 
contributing to the UCN source project at 
the Paul Scherrer Institute who are indispensable for the realization of this project. 
Support of our colleagues at PF2 - ILL and the Mainz TRIGA UCN 
Source during component testing
and of Jagellonian University Cracow and LPSC Grenoble 
is gratefully acknowledged. 
PNPI contributed in the early planning of the project.

\end{document}